# DYNAMICS OF HUMAN SOCIAL NETWORKS PEOPLE, TIME, RELATIONSHIPS AND PLACES

Rahman O. Oloritun
Masdar Institute
P. O Box 54224
Abu Dhabi, UAE
+971 5675354283
roloritun@masdar.ac.ae

Alex P. Pentland
Massachusetts Institute of Technology
E15-387
20 Ames Street, Cambridge, MA 02139
+1 6172530648
sandy@media.mit.edu

Inas Khayal
Masdar Institute and MIT
P. O Box 54224
Abu Dhabi, UAE
+971 2 810 9128
ikhayal@masdar.ac.ae

## ABSTRACT

The availability of advanced social interaction sensing technologies provides fine grained data for social network analysis. Although traditional methods of gathering social network data may be subject to human ability to recall social details, people's rating of their closeness to persons in their network is important.

This study assesses the relationship amidst closeness ratings, sensed interactions and shared places of recreation. The study found that people tend to give high closeness ratings to people with whom they spend more time. Shared places are correlated to social closeness ratings but not to length of interactions. The results of this study highlight the importance of sensed interactions and closeness ratings.

## INTRODUCTION

There are several intuitions about the human social network. Some of these intuitions or generally accepted assumptions suggest that domestic partners have high durations of interactions. People who spend more time together and couples are expected to have high closeness ratings. Another common belief is that people who are socially close or are in a domestic partnership tend to share places of recreation.

This paper shows the relationships and interactions among the sensed networks (based on measures of time spent in proximity by an interacting pair), social closeness ratings, domestic partnerships, and shared places.

Several studies have emphasized the abundance of information present in networks. The data from networks have been used to study diffusion of infections (Stehlé, Voirin, Barrat, Cattuto, Colizza, Isella, et al., 2011) happiness (Fowler and Christakis, 2008), information (Panisson, Barrat, Cattuto, Van den Broeck, Ruffo, and Schifanella, 2011) and obesity (Christakis and Fowler, 2007). These studies have shown that social ties affect our lives in several ways ranging from life choices to health.

Past studies on human social networks were affected by human ability to recall exact duration of social interactions, the exact numbers and names of people with whom they interacted. However, individuals are expected to be quite precise when they rate their social nearness to friends and other individuals in their network. The major problem is recalling everyone with whom there was an interaction.

The availability of new data acquisition techniques for logging human face-to-face interaction is opening new avenues for understanding the dynamics of interactions in social networks and providing researchers with access to almost complete social networks. Time-resolved face-to-face interactions by individuals in real-world settings can be captured using embedded sensing techniques. Interaction sensing devices have been used in several studies. Some of these studies (T. Choudhury and Pentland, 2004; Tanzeem Choudhury, Philipose, Wyatt, and Lester, 2006) analyzed face-to-face interactions captured via device called the Sociometer. The device records when and if people were conversing. RFIDs (Isella, Stehle, Barrat, Cattuto, Pinton, and den Broeck, 2011; Panisson, et al., 2011; Stehlé, et al., 2011), and mobile phones (Abdelzaher, Anokwa, Boda, Burke, Estrin, Guibas, et al., 2007; Eagle and Pentland, 2006; Eagle, Pentland, and Lazer, 2009; Miluzzo, Cornelius, Ramaswamy, Choudhury, Liu, and Campbell, 2010) have also been used to sense interaction networks.

## METHODLOGY

This section presents the social network analysis approach used in this paper to evaluate the relationships that within people, their social ties (closeness, domestic partnerships and lengths of interaction) and places of recreation. The approach consists of network/graph construction, visualization of the networks, network correlations and regressions.

## Study Population

The participants in this study were members of a young-family residential living community adjacent to a university in North America. All members of the community are couples, and at least one the members is affiliated with the university, usually as a graduate student.

The entire community is composed of over 400 residents, approximately half of which have children, with low- to mid-range household income. The residence has a vibrant community life, with many ties of friendship between its members. The data used for this analysis was collected from 101 participants composed of 55% males and 45% females.

## Data Collection

Participants were provided Android smartphones with an open source software sensing platform (FunF) that allowed for the detection of social interactions via Bluetooth proximity sensing. A mobile enabled online survey was also made available to participants in order to capture details like social closeness ratings (on a Likert Scale (0[least]-8[highest])), domestic partners and other behaviors of interest. In the survey, participants were asked to exhaustively choose the locations where they exercise from a list of places provided. The list of places included six popular recreation locations, the individuals' apartment and others (for locations not commonly used or very far from the study site).

The FunF application and data collected are described in detail in study titled Social fMRI: Investigating and shaping social mechanisms in the real world by Aharony, Pan, Ip, Khayal, and Pentland (2011). The data collection platform is shown in Figure 1. The data used for this study is found online at http://realitycommons.media.mit.edu/friendsdataset.html, with an appraisal of the data quality, noise levels and sensor characteristics. The data used in this study spans a period of 3 months (October to December 2010), approximately 13 weeks.

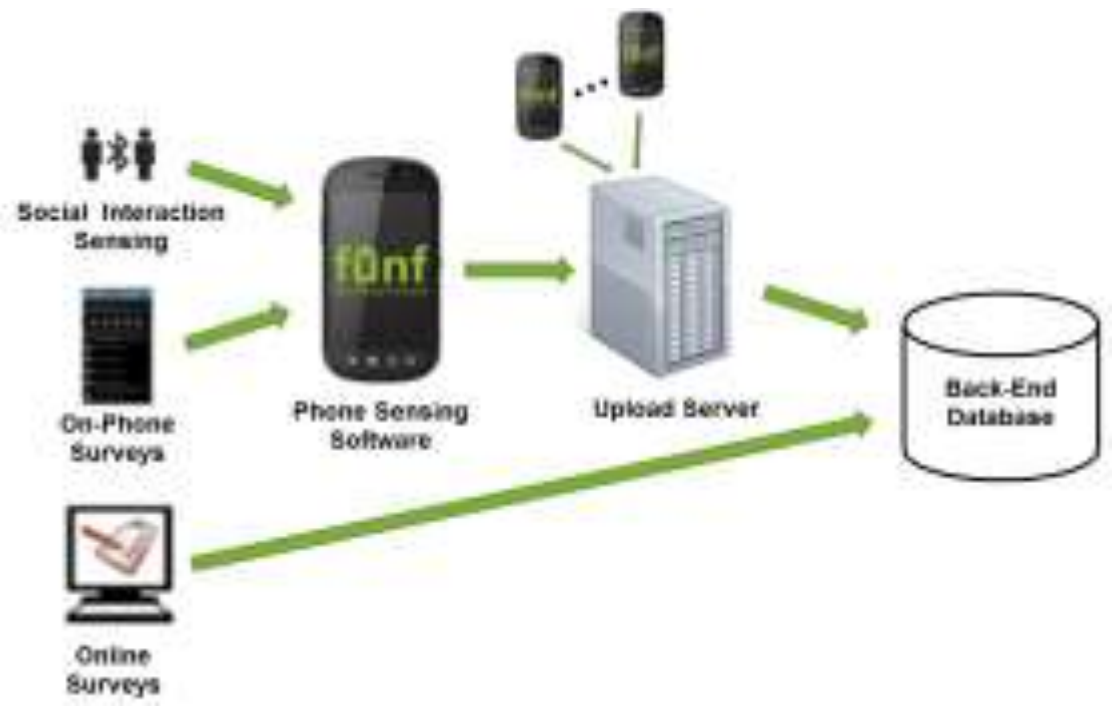

*Figure 1: The data collection platform*

## Analysis

The interaction network was aggregated to two (2) levels, week and for the entire three months period. The week level consists of 13 interaction networks (one for each week) where the edge weight is the duration of interaction. Only interactions networks were available on the weekly time scale. Five (5) networks were constructed from the data collected. The networks include (1) bipartite network of participants and places of recreation, (2) a network of interaction with the hours of interaction per week as edge weights, (3) closeness ratings network and (4) shared places. The fifth network is obtained via the projection of the bipartite network of participants and places of recreation from a two-way network to a one-mode network of participants where the edge weight is the number of shared places of recreation.

### *Networks Aggregation and Projection*

The social closeness ratings were collected at the start of the experiment and end of the three months period. The reported social closeness was averaged to obtain the weights for the social closeness ratings network.

In order to compensate for participants without interactions measures for some of the weeks, the total time for each dyad was divided by the total number of weeks. Thus, the duration per week per dyad was obtained and used as the weights of the interaction network.

Let $\boldsymbol{W_i}$ represent the networks for week $i$ where $i = 1, 2 \ldots 13$ and the **A** be the aggregated interaction network, then the aggregate interaction network **A** is obtained by the following equation;

$$A = \frac{1}{13}\sum_{i=1}^{13} W_i$$

### *Nodal Level Analysis*

The nodal level properties of each of the weekly interaction networks were compared with that of other weeks' interaction using Spearman Rank correlation. Shapiro-Wilk's (Patrick, 1982) test of normality showed that the nodal properties are not normally distributed. The nodal level properties evaluated consist of centrality measures such as weighted degree, betweenness and closeness scores. Correlation of the centrality measures from the weeks' interaction networks provides insights into how the succeeding nodal level properties can be inferred from the preceding.

The nodal level properties of aggregated interaction, social closeness ratings, domestic partnerships and shared places networks were also correlated to assess how the centrality of participants changes across the networks.

Correlation of centrality measures is used to evaluate the relationship between similar centrality measures of two networks. The correlation can be used to infer if a node will likely have the same role (determined by a centrality measure) in a network in another network.

### *Network Level Analysis*

QAP (Quadratic Assignment Procedure) correlations were used to examine the relationship between the networks and QAP regressions to understand how a combination of other networks explains a network. The quadratic assignment procedure (QAP) is an approach for statistical significance testing of social network data. An assumption of parametric statistical techniques, which determine statistical significance by comparing observed values to appropriate theoretical distributions, is that the observations being analyzed are independent of one another. This assumption is not accurate and does not hold in social network analysis.

QAP is a non-parametric technique, meaning it does not rely on assumptions of independence; it is also a general procedure that is frequently used for both correlation and multiple regression analysis. The QAP approach (Krackhardt, 1988) estimates regression model coefficients and then uses random permutations of the network data to generate a distribution of coefficient estimates from random networks with the same structure. The actual estimates are then compared with this generated distribution to test for significance. Improvements in this procedure have been made to ensure conservative estimation of standard errors across less-than-ideally structured data (Dekker, Krackhardt, and Snijders, 2007).

Thirteen weekly interaction networks were correlated to assess if the interaction networks are serially correlated. The aggregated interaction, social closeness ratings, domestic partnerships and shared places networks were correlated to evaluate the relationship between these human social networks. Three (3) QAP models were developed to evaluate the relationships among the networks.

Model 1 has the interaction network as the dependent variable with social closeness ratings, domestic partnerships and shared places networks, respectively, as independent variables.

Model 2 has the social closeness ratings network as the dependent variable with interaction, domestic partnerships and shared places networks, respectively, as independent variables while model 3 has the shared places network as the dependent variable with social closeness ratings, domestic partnerships and interaction networks, respectively, as independent variables.

## RESULTS

The results of analysis performed at the nodal level and network level are presented in this section.

## Nodal Level Analysis Results

### *Weekly interaction networks*

Figure 2 depicts the spearman rank correlation matrix of the weighted degrees of the 13 weekly interaction networks. It can be inferred from Figure 2 that $WD_n$ and $WD_{n+1}$ are have high positive correlation coefficients where $WD_n$ is the weighted degree for weekly interaction networks $n$ $(n=1,2,3,\ldots,13)$. This suggests that the weighted degree of a week's interaction network can be inferred from preceding week's interaction network.
The correlations of other centralities measures (betweenness and closeness) for the weekly networks did not show the serial correlation pattern noticed in the correlation of the weighted degrees.

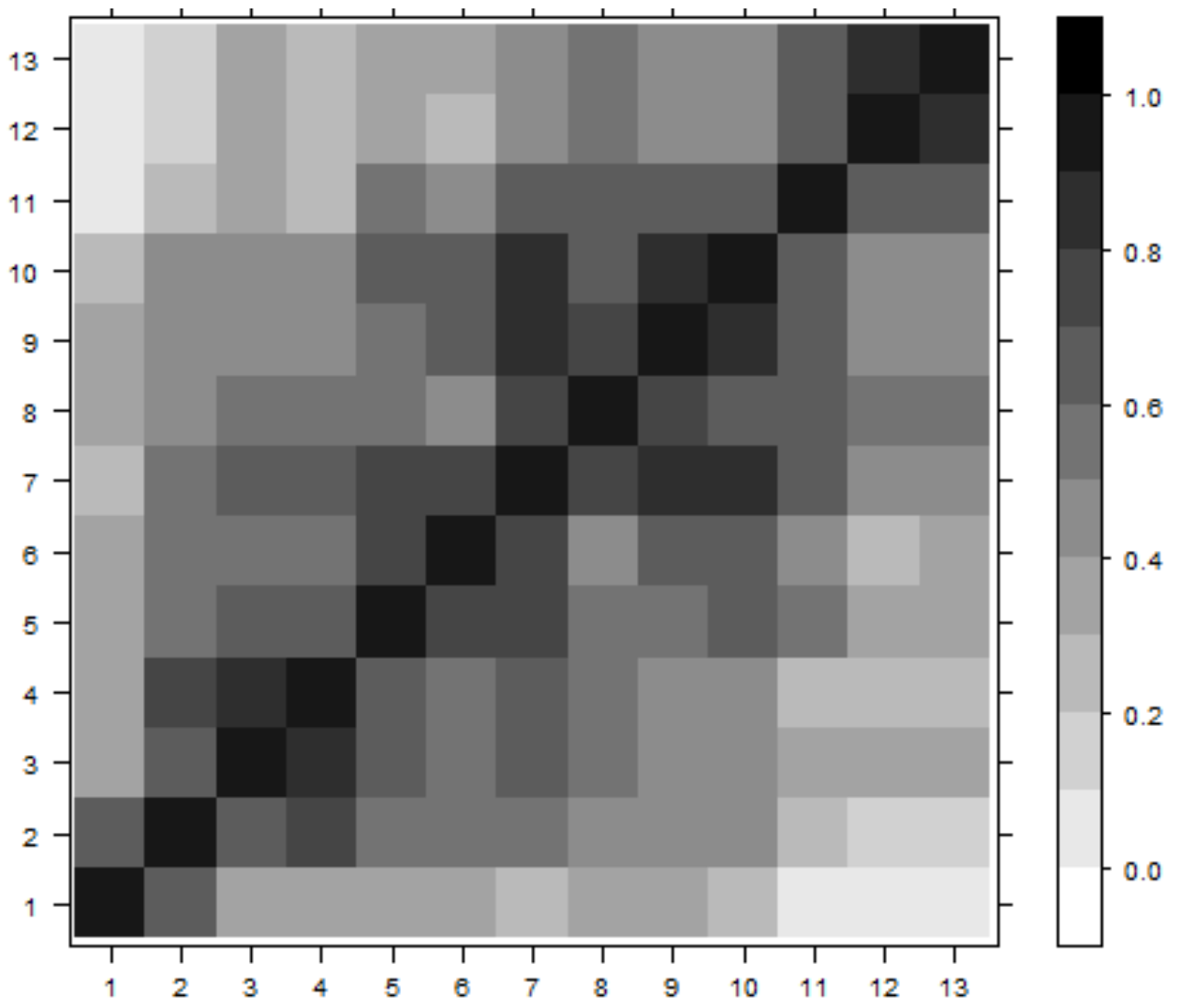

*Figure 2: Spearman Rank Correlation Matrix the weighted degree of the weekly (13 weeks) interaction networks.*

### *Aggregated Networks*

The correlation matrix of the centralities measures (weighted degree, betweenness and closeness) is depicted in Figure 3. *WD1*, *WD2*, *WD3*, and *WD4* are the weighted degree for the aggregated interaction, social closeness ratings, domestic partnerships, and shared places networks, respectively. *B1*, *B2*, *B3*, and *B4* are the betweenness scores for the aggregated interaction, social closeness ratings, domestic partnerships, and shared places networks, respectively; while *C1*, *C2*, *C3*, and *C4* are the closeness scores for the aggregated interaction, social closeness ratings, domestic partnerships, and shared places networks, respectively.

Although all the centrality measures are placed in a single correlation matrix, this study only reports correlation of similar centrality measures i.e. degrees are correlate with degrees, betweenness with betweenness and closeness with closeness. This study only reports correlation values greater than 0.4.

*WD1*, the weighted degree of the aggregated interaction network, and *WD2*, the weighted degree of the self-rated closeness network are correlated with coefficient of 0.45 with p-value < 0.01. The closeness centrality measures from the aggregated interaction network, *C1* and closeness centrality measures from the self-rated closeness network, *C2* were found to be correlated with high correlation coefficient of 0.52 *(p-value < 0.01)*.

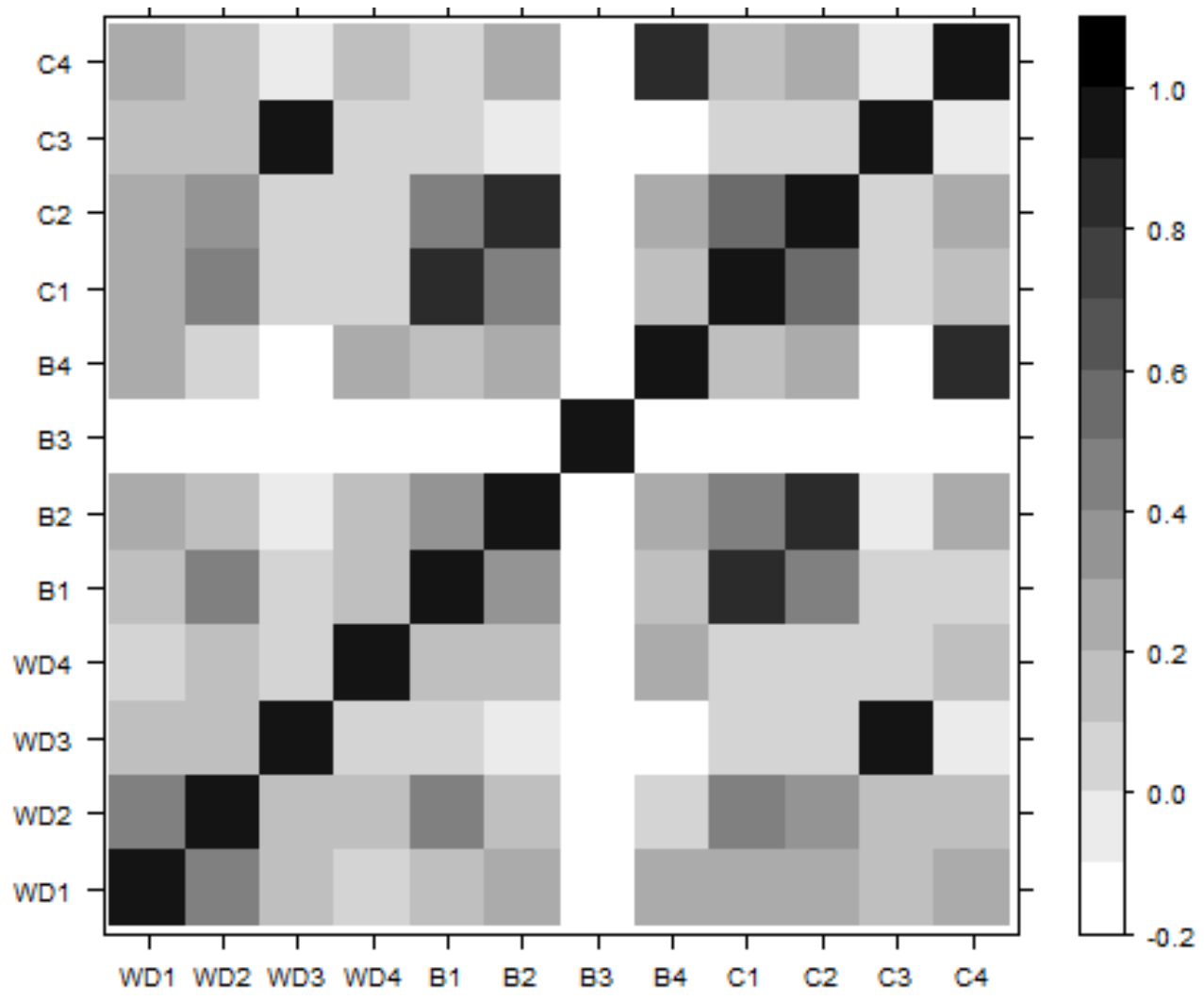

*Figure 3: Spearman Rank Correlation Matrix of networks centrality measures. WD1, WD2, WD3, WD4 are the weighted degree, and B1, B2, B3, B4 are the betweenness centrality measures while C1, C2, C3, C4 are the closeness centrality measures for the aggregated interaction, social closeness ratings, domestic partnerships and shared places networks, respectively.*

<u>**Network Level Analysis Results**</u>

***Weekly interaction networks***

Let $\boldsymbol{W_i}$ represent the networks for week $i$ where $i = 1, 2 \ldots 13$. It can be inferred from correlation matrix of the weekly interaction networks shown in Figure 4 that the 13 networks are serially correlated such $\boldsymbol{W_i}$ and $\boldsymbol{W_{i+1}}$ have positive correlation coefficient close to 1. This suggests that the interaction network of the succeeding week can be inferred from the preceding week interaction network. This is evidenced by the darker squares around the black diagonal.

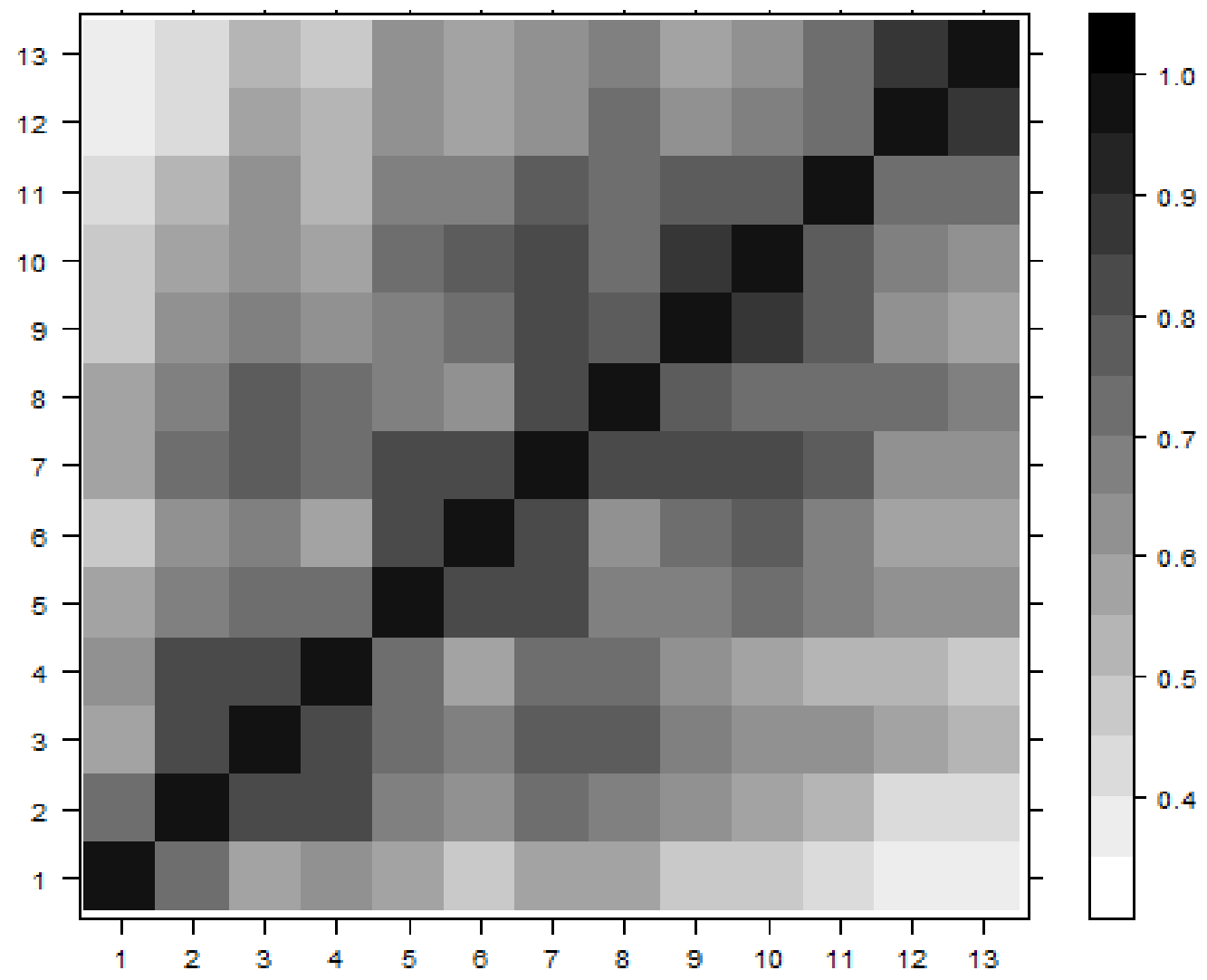

*Figure 4: QAP correlation matrix of weekly (13 weeks) interactions networks*

***Aggregated Networks***

There are six major places of interest in this study and 101 participants. The size of the nodes in Figure 5 represents the degree of the node with red (big) nodes as places and blue nodes represent people. There are people who do not use any of the shared places and they appear as unconnected vertices in Figure 5. They reported either exercising in their apartments or at locations that less common or very far from the community. Figure 6 depicts a one-mode network that is obtained from the bipartite network in Figure 5. The blue vertices in Figure 6 represent male participants while red vertices represent female participants. Figure 5 and 6 both have 11 isolates i.e. vertices without any edge incident on them.

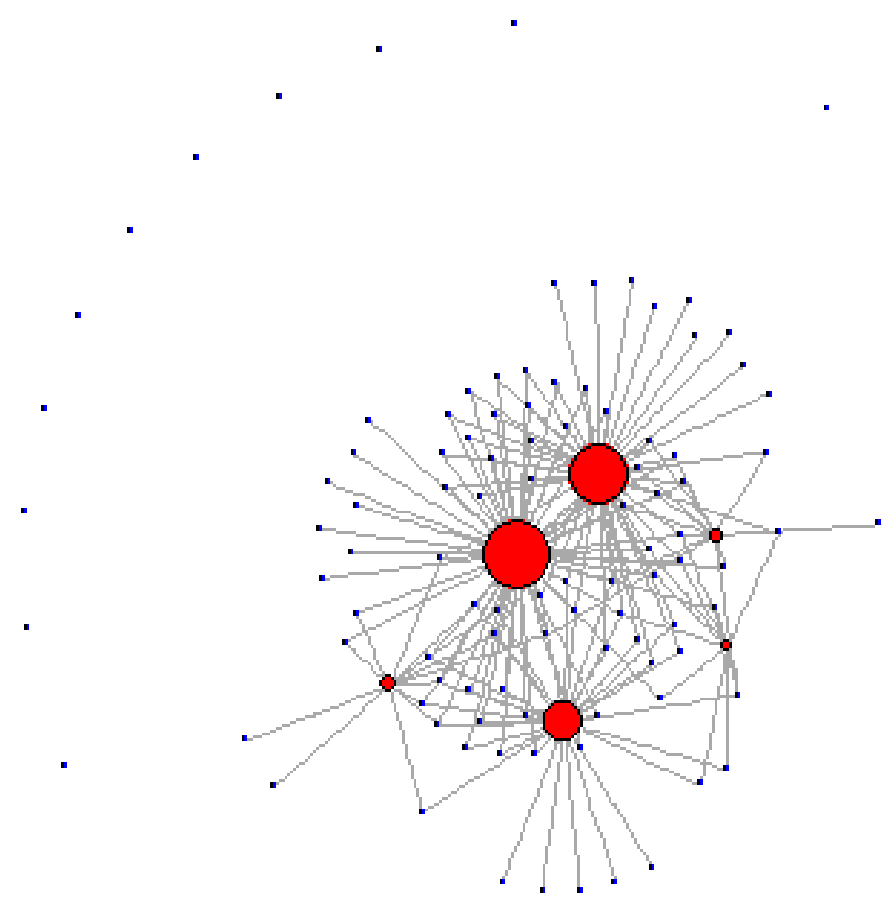

*Figure 5: Bipartite Graph of participants and places. Red vertices are places and blue vertices are people.*

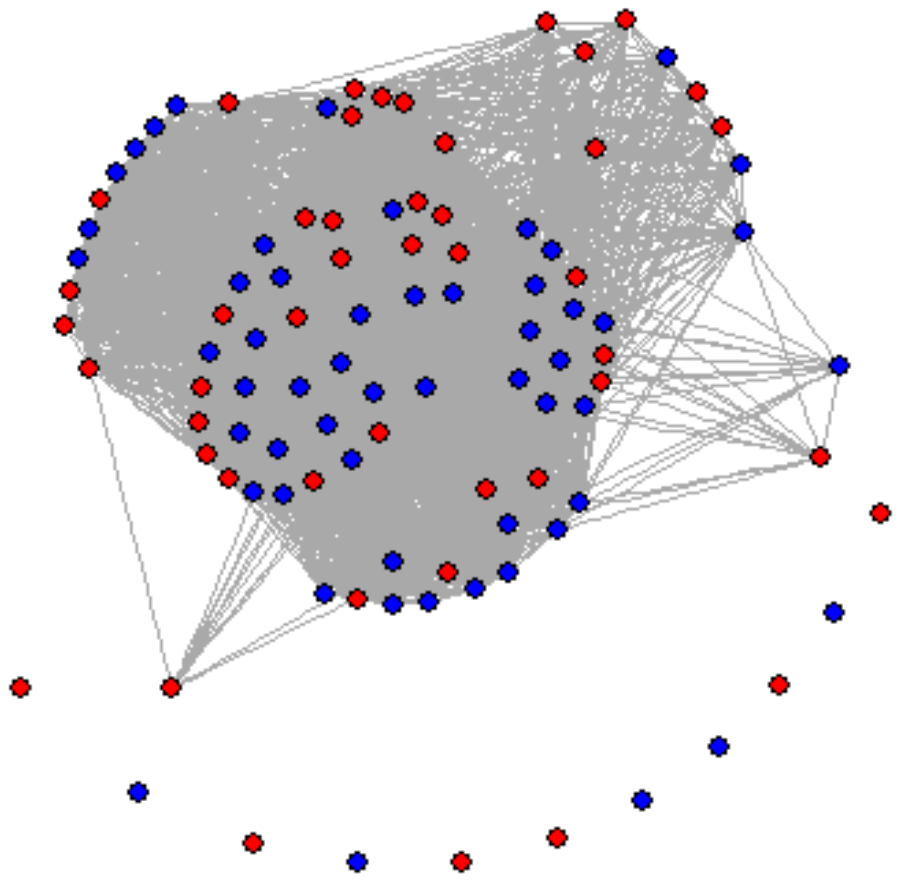

*Figure 6: Shared places network. Blue Vertices are Males and Red vertices females.*

Figure 7 depicts the correlation matrix of 4 networks (1: Interaction Network; 2: Social closeness ratings network; 3: Domestic partnerships; 4: Shared Places Network). Domestic partnerships and interaction networks had the highest correlation of 0.57*(p-value<0.0001).* The next highest correlation of 0.35*(p-value<0.0001)* was between interaction and social closeness ratings network. A correlation of 0.28*(p-value<0.0001)* exists between the domestic partnerships and social closeness ratings networks. Shared places and social closeness ratings networks had correlation value of 0.08*(p-value<0.05).*

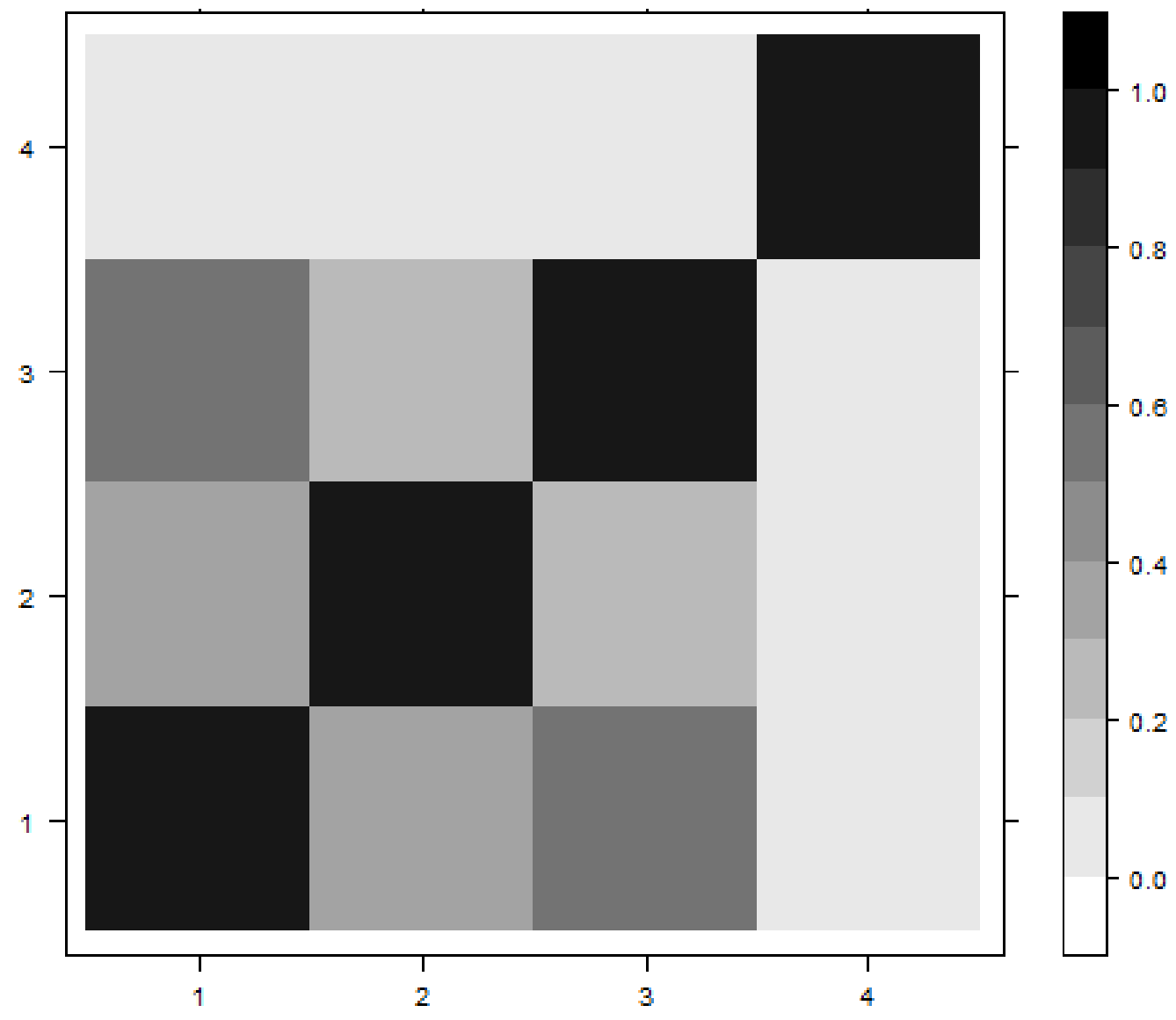

*Figure 7: QAP Correlation Matrix of Networks. 1: Aggregated Interaction Network; 2: Social closeness ratings network; 3: Domestic partnerships; 4: Shared Places Network.*

Burris (2005) argues that when interpreting QAP regression results, the focus should be on the comparative magnitude of the coefficients, rather than on the overall model $R^2$ or the level of statistical significance for each coefficient. Discussion of results shown in Table I will focus on the comparative magnitude of those coefficients, which are significant at $p < 0.1$.

*Table 1: Regression Models for Networks*

| | **Model 1** | **Model 2** | **Model 3** |
|---|---|---|---|
| | ***Interaction*** | ***Social closeness ratings*** | ***Shared Places*** |
| | ***Estimates (β)*** | ***Estimates (β)*** | ***Estimates (β)*** |
| Intercept | 0.047 | 0.870* | 0.749* |
| Interactions | | 0.196* | 0.002 |
| Social closeness ratings | 0.336* | | 0.032+ |
| Domestic partnerships | 15.976* | 2.089* | 0.109 |
| Shared Places | 0.021 | 0.166+ | |
| Adjusted $R^2$ | 0.373* | 0.142* | 0.007* |

*+ p <0.1, * p <0.0001*

In model 1, where the dependent variable is the interaction network, the estimates of the social closeness ratings and domestic partnerships networks are very significantly different from zero (0).

In model 2, where the dependent variable is the social closeness ratings network, the estimates of the interaction and domestic partnerships networks are significantly different from zero (0). The estimate for shared places in model 2 is also significantly different for zero with a p-value of *0.089*.

In model 3, where the dependent variable is the shared places network, the estimate of the social closeness ratings network is significantly different *(p=0.077)* from zero (0).

## DISCUSSION

This study explores general intuitions about the relationships that exist among duration of interactions, domestic partnerships, reported level of closeness and shared places.

Most studies rely on surveys that ask participants to list people with whom they are friends, acquaintances or interact. Social network data collected via survey is subject to the human ability to precisely recall all the people they interact with or the length of time they spent interacting. The findings of this study suggest that people tend to interpret closeness in terms of length of time of social interactions. People tend to report that they are closer to people with whom they spend more time interacting. This is inferred from the positive correlation between duration of interactions and the social closeness ratings network.

The intuition that domestic partners (couples) spend most times together is explicitly implied by the high positive correlation of the duration of interactions and the domestic partnership network.

Social closeness ratings network is the only network that exhibits a relationship with the shared places network. This implies that peoples rating of how close they are to someone is a stronger pull to exercise at same places than the overall time they spend together. This suggests that futures studies on human social networks will do well to combine new social interaction sensing technologies and people's perception of their social relationships.

The lack of a wide range of ages in the study population may limit the generality of the study although the study population includes subjects with varying levels of income, and from different cultures. Another limitation of this study is the assumption that interaction occurs whenever the socially aware phones (devices used to capture the interaction data analyzed in this study) are within their Bluetooth transceivers range even if the participants do not engage in any form of observable interactions. It should be noted that the range of the sensor devices is generally visible to participants, i.e., participants will most of the time see the other participants if their sensors can connect.

## CONCLUSION

This study explored the relationships among several dynamics and drivers of real world human social networks such as duration of interactions, relationships (social nearness and domestic partnerships) and public spaces. The study found that people tend be closer to people with whom they spend more time interacting. Shared places are correlated to social closeness ratings. This study suggests that personal rating of social near-ness (closeness) is important in the study human social network.

This study also highlighted the importance of social interactions measured using embedded sensing techniques in the real world, presenting a new avenue for understanding individual behavior and explanation of various social ties. These results paint a bright picture for future studies of social networks fueled by the latest advances in wireless communication and embedded sensing technologies.

## APPENDIX

*Table 2: Spearman Rank Correlation Matrix the weighted degree of the weekly (13 weeks) interaction networks.*
**: p<0.05, **: p<0.01*

| | **1** | **2** | **3** | **4** | **5** | **6** | **7** | **8** | **9** | **10** | **11** | **12** | **13** |
|---|---|---|---|---|---|---|---|---|---|---|---|---|---|
| **1** | | 0.62** | 0.35** | 0.36** | 0.35** | 0.37** | 0.29** | 0.31** | 0.30** | 0.26** | 0.10 | 0.02 | 0.07 |
| **2** | 0.62** | | 0.69** | 0.70** | 0.54** | 0.52** | 0.55** | 0.46** | 0.49** | 0.42** | 0.22* | 0.11 | 0.15 |
| **3** | 0.35** | 0.69** | | 0.80** | 0.63** | 0.59** | 0.64** | 0.57** | 0.47** | 0.49** | 0.39** | 0.36** | 0.37** |
| **4** | 0.36** | 0.70** | 0.80** | | 0.66** | 0.53** | 0.62** | 0.51** | 0.45** | 0.46** | 0.30** | 0.27** | 0.26** |
| **5** | 0.35** | 0.54** | 0.63** | 0.66** | | 0.76** | 0.72** | 0.57** | 0.59** | 0.62** | 0.53** | 0.35** | 0.37** |
| **6** | 0.37** | 0.52** | 0.59** | 0.53** | 0.76** | | 0.70** | 0.45** | 0.63** | 0.69** | 0.48** | 0.27** | 0.38** |
| **7** | 0.29** | 0.55** | 0.64** | 0.62** | 0.72** | 0.70** | | 0.75** | 0.81** | 0.82** | 0.68** | 0.43** | 0.44** |
| **8** | 0.31** | 0.46** | 0.57** | 0.51** | 0.57** | 0.45** | 0.75** | | 0.73** | 0.70** | 0.61** | 0.55** | 0.51** |
| **9** | 0.30** | 0.49** | 0.47** | 0.45** | 0.59** | 0.63** | 0.81** | 0.73** | | 0.87** | 0.63** | 0.42** | 0.41** |
| **10** | 0.26** | 0.42** | 0.49** | 0.46** | 0.62** | 0.69** | 0.82** | 0.70** | 0.87** | | 0.69** | 0.42** | 0.47** |
| **11** | 0.10 | 0.22* | 0.39** | 0.30** | 0.53** | 0.48** | 0.68** | 0.61** | 0.63** | 0.69** | | 0.64** | 0.63** |
| **12** | 0.02 | 0.11 | 0.36** | 0.27** | 0.35** | 0.27** | 0.43** | 0.55** | 0.42** | 0.42** | 0.64** | | 0.85** |
| **13** | 0.07 | 0.15 | 0.37** | 0.26** | 0.37** | 0.38** | 0.44** | 0.51** | 0.41** | 0.47** | 0.63** | 0.85** | |

*Table 3: Spearman Rank Correlation Matrix of the networks' centrality measures. WD1, WD2, WD3, WD4 are the weighted degree, and B1, B2, B3, B4 are the "betweenness" centrality measures while C1, C2, C3, C4 are the closeness centrality measures for the aggregated interaction, social closeness ratings, domestic partnerships and shared places networks, respectively. B3 has all values as zeros. *: p<0.05, **: p<0.01*

| | **WD1** | **WD2** | **WD3** | **WD4** | **B1** | **B2** | **B4** | **C1** | **C2** | **C3** | **C4** |
|---|---|---|---|---|---|---|---|---|---|---|---|
| **WD1** | | **0.45**** | **0.18** | **0.10** | 0.12 | 0.22* | 0.22* | 0.21* | 0.29** | 0.18 | 0.22* |
| **WD2** | **0.45**** | | **0.17** | **0.14** | 0.49** | 0.19 | 0.09 | 0.48** | 0.37** | 0.17 | 0.11 |
| **WD3** | **0.18** | **0.17** | | **0.00** | 0.03 | -0.03 | -0.10 | 0.09 | 0.08 | 1.00** | -0.06 |
| **WD4** | **0.10** | **0.14** | **0.00** | | 0.18 | 0.11 | 0.25* | 0.08 | 0.00 | 0.00 | 0.11 |
| **B1** | 0.12 | 0.49** | 0.03 | 0.18 | | **0.37**** | **0.10** | 0.85** | 0.43** | 0.03 | 0.08 |
| **B2** | 0.22* | 0.19 | -0.03 | 0.11 | **0.37**** | | **0.27**** | 0.41** | 0.82** | -0.03 | 0.25* |
| **B4** | 0.22* | 0.09 | -0.10 | 0.25* | **0.10** | **0.27**** | | 0.18 | 0.25* | -0.10 | 0.84** |
| **C1** | 0.21* | 0.48** | 0.09 | 0.08 | 0.85** | 0.41** | 0.18 | | **0.52**** | **0.09** | **0.17** |
| **C2** | 0.29** | 0.37** | 0.08 | 0.00 | 0.43** | 0.82** | 0.25* | **0.52**** | | **0.08** | **0.26**** |
| **C3** | 0.18 | 0.17 | 1.00** | 0.00 | 0.03 | -0.03 | -0.10 | **0.09** | **0.08** | | **-0.06** |
| **C4** | 0.22* | 0.11 | -0.06 | 0.11 | 0.08 | 0.25* | 0.84** | **0.17** | **0.26**** | **-0.06** | |

*Table 4: QAP correlation matrix of weekly (13 weeks) interactions networks (all at p<0.01)*

| | **1** | **2** | **3** | **4** | **5** | **6** | **7** | **8** | **9** | **10** | **11** | **12** | **13** |
|---|---|---|---|---|---|---|---|---|---|---|---|---|---|
| **1** | | 0.71 | 0.57 | 0.54 | 0.45 | 0.38 | 0.5 | 0.56 | 0.48 | 0.4 | 0.31 | 0.2 | 0.15 |
| **2** | 0.71 | | 0.78 | 0.74 | 0.63 | 0.56 | 0.69 | 0.66 | 0.65 | 0.56 | 0.45 | 0.33 | 0.25 |
| **3** | 0.57 | 0.78 | | 0.86 | 0.73 | 0.65 | 0.77 | 0.76 | 0.69 | 0.65 | 0.6 | 0.55 | 0.46 |
| **4** | 0.54 | 0.74 | 0.86 | | 0.69 | 0.58 | 0.76 | 0.74 | 0.62 | 0.62 | 0.5 | 0.47 | 0.39 |
| **5** | 0.45 | 0.63 | 0.73 | 0.69 | | 0.81 | 0.79 | 0.65 | 0.65 | 0.7 | 0.65 | 0.55 | 0.54 |
| **6** | 0.38 | 0.56 | 0.65 | 0.58 | 0.81 | | 0.81 | 0.56 | 0.7 | 0.76 | 0.66 | 0.46 | 0.5 |
| **7** | 0.5 | 0.69 | 0.77 | 0.76 | 0.79 | 0.81 | | 0.82 | 0.82 | 0.84 | 0.76 | 0.58 | 0.54 |
| **8** | 0.56 | 0.66 | 0.76 | 0.74 | 0.65 | 0.56 | 0.82 | | 0.77 | 0.7 | 0.65 | 0.64 | 0.54 |
| **9** | 0.48 | 0.65 | 0.69 | 0.62 | 0.65 | 0.7 | 0.82 | 0.77 | | 0.88 | 0.75 | 0.53 | 0.48 |
| **10** | 0.4 | 0.56 | 0.65 | 0.62 | 0.7 | 0.76 | 0.84 | 0.7 | 0.88 | | 0.8 | 0.56 | 0.54 |
| **11** | 0.31 | 0.45 | 0.6 | 0.5 | 0.65 | 0.66 | 0.76 | 0.65 | 0.75 | 0.8 | | 0.66 | 0.66 |
| **12** | 0.2 | 0.33 | 0.55 | 0.47 | 0.55 | 0.46 | 0.58 | 0.64 | 0.53 | 0.56 | 0.66 | | 0.85 |
| **13** | 0.15 | 0.25 | 0.46 | 0.39 | 0.54 | 0.5 | 0.54 | 0.54 | 0.48 | 0.54 | 0.66 | 0.85 | |

*Table 5: QAP Correlation Matrix of Networks. 1: Aggregated Interaction Network; 2: Social closeness ratings network; 3: Domestic partnerships; 4: Shared Places Network. (*: p<0.05, **: p<0.01, ***: p<0.00001).*

| | **1** | **2** | **3** | **4** |
|---|---|---|---|---|
| 1 | | 0.36*** | 0.57*** | 0.04** |
| 2 | 0.36*** | | 0.28*** | 0.08* |
| 3 | 0.57*** | 0.28*** | | 0.04*** |
| 4 | 0.04** | 0.08* | 0.04*** | |